\documentclass[onecollarge,natbib]{svjour2}
\bibpunct{[}{]}{,}{n}{}{,} 
\smartqed  
\usepackage{graphicx}
\usepackage{mathptmx}      
\journalname{Few-Body Systems}
\begin{document}

\title{\boldmath
Are hyperon resonances required in the 
elementary $K^+\Lambda$ photoproduction?\footnote{Dedicated to
Professor Henryk Witala at the occasion of his 60th birthday.}}

\titlerunning{Are hyperon resonances really required?}

\author{T. Mart and N. Nurhadiansyah}


\institute{T. Mart \at
              Departemen Fisika, FMIPA, Universitas Indonesia,
              Depok 16424, Indonesia\\
              \email{tmart@fisika.ui.ac.id}
}

\date{Received: date / Accepted: date}

\maketitle

\begin{abstract}
  We have investigated the role of hyperon resonances in the
  kaon photoproduction process, $\gamma p\to K^+\Lambda$, by using
  a covariant isobar model. To this end, new experimental 
  data are used in the fitting process, whereas the old
  SAPHIR 1998 data are also used for comparison. The result indicates 
  that the 
  $\Lambda(1600)P_{01}$ and $\Lambda(1810)P_{01}$ hyperon resonances
  can significantly reduce the $\chi^2$ and, simultaneously, can 
  increase the hadronic form factor cut-off in the background
  terms. This finding is different from the result of the 
  previous studies, which showed that the 
  $\Lambda(1800)S_{01}$ was important for this purpose, instead of
  the $\Lambda(1600)P_{01}$.
\end{abstract}

\section{Introduction}
\label{intro}
Recently, it has been shown that kaon photoproduction process provides
an important tool for investigating missing resonances, 
the resonances predicted by constituent quark models but not 
observed yet by the  Particle Data Group (PDG) \cite{pdg}.
The relatively small decay widths to pion-nucleon channel
are believed to be the reason behind this fact. However, certain
constituent quark models \cite{NRQM,capstick94} predict 
that these resonances have larger decay widths to other 
channels, such as kaon-hyperon ones. Thus, kaon photoproduction
is well suited to shed more light on the existence of these
missing resonances.

Whereas nucleon resonances have been intensively investigated
by using kaon photoproduction, there has been no thorough study on
hyperon resonances by means of this process. Ideally, such
a study should be performed by using the kaon-nucleon scattering,
in which hyperon resonances propagate in the $s$-channel. 
Furthermore, kaon-nucleon scattering experiment has become 
the main focus of recent
experimental activities at JPARC in Japan \cite{noumi}. 
Nevertheless, in order to achieve an accuracy  similar to
that obtained in experiments using electromagnetic
beams performed at Jefferson Lab, MAMI, and Spring8,
one needs very intense kaon beam. Moreover, it is widely 
understood that hadronic interaction is not as ''clean''
as the electromagnetic one. In view of this, kaon
photoproduction could offer a complementary solution, since the 
hyperon resonance propagates in the $u$-channel and, therefore, 
the process is sensitive to certain observables at backward kaon 
scattering angles.

More than a decade ago it was found that the second peak in the 
cross section of SAPHIR data \cite{saphir98} can be nicely 
reproduced by including a missing $D_{13}(1895)$ nucleon resonance 
in a covariant isobar model. \cite{missing-d13}. An isobar model 
which includes this resonance and hadronic form factors
was fitted to the SAPHIR data. The form factor cut-off $\Lambda$
was allowed to
vary during the fitting process and it was found that $\Lambda=0.64$ GeV. 
Although there is no tool to measure this form factor directly, 
since it is an off-shell form factor, such a small cut-off was considered 
to be unreasonably soft \cite{janssen} and the form factors
were considered as an artificial mechanism required just to suppress
the diverging Born terms at high energies. Instead of using this
resonance and an artificial cut-off, Ref.~\cite{janssen} proposed
to use $\Lambda(1800)S_{01}$ and $\Lambda(1810)P_{01}$, which were
claimed to have the same ability to suppress the Born terms, 
while simultaneously to keep the form factor sufficiently ``hard''.

Since the number of experimental data at present is almost one
order of magnitude larger than the number when the claim was
proposed, we believe that it is timely to check whether or not 
the  $\Lambda(1800)S_{01}$ and $\Lambda(1810)P_{01}$ hyperon 
resonances are still able to help the model 
in improving the agreement with experimental data,
while keeping the hadronic cut-off reasonably ``hard''. 
It is the main purpose of 
the present paper to discuss the limitation of the claim. Furthermore,
it is also important to investigate the role of other hyperon
resonances in kaon photoproduction. Since at present 
the number of data becomes 
sufficiently large, while the corresponding error bars are
considerably small, we believe that there are just few 
degrees of freedom left in the model. This is obviously in contrast
to the former situation.

In Section \ref{sec:model} we briefly review the isobar model used
in our investigation. Section \ref{sec:result} discusses the numerical
results of our investigation. We will summarize our discussion and
conclude our findings in Section \ref{sec:conclusion}.

\section{Isobar Model}
\label{sec:model}
\begin{figure}[t]
\centering
  \includegraphics[width=0.9\textwidth]{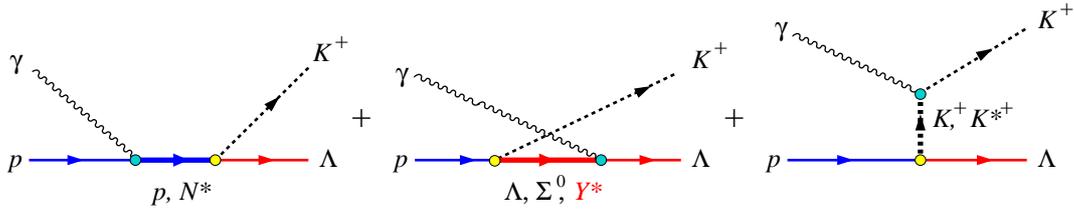}
\caption{Feynman diagrams for $K^+\Lambda$ photoproduction
  on the proton. The hyperon resonance $Y^*$ contributes to the
  $u$-channel process (middle diagram).}
\label{fig:feynman}
\end{figure}

The elementary process for $K^+\Lambda$ photoproduction
on the proton target can be written as
\begin{eqnarray}
  \gamma (k)+ p(p_p) \to K^+ (q) + \Lambda (p_\Lambda)~.
\end{eqnarray}
Theoretically, this process can be described by utilizing 
an isobar model. At a tree-level approximation, the 
process is schematically shown in Fig.~\ref{fig:feynman}. 
The elementary 
transition operator can be calculated from the appropriate
Feynman diagrams in Fig.~\ref{fig:feynman} and 
decomposed into
\begin{eqnarray}
  {\cal M}_{\rm fi} = 
  \bar{u}({\bf p}_\Lambda)\,\sum_{i=1}^{4}\, 
  A_i(s,t,u)\,M_i\, u({\bf p}_p) ~.
\label{elementary_tr_op}
\end{eqnarray}
where $s,t$ and $u$ are the Mandelstam variables. 
The gauge and Lorentz invariant matrices $M_{i}$ can be 
found, e.g., in Ref.~\cite{mart_hyp}. All relevant observables
required in this investigation can be calculated from the 
functions $A_i(s,t,u)$ (see, e.g. Ref.~\cite{knochlein}).

The background terms of the model consist of the standard 
$s$-, $t$-, and $u$-channel Born terms and appropriate kaon and
hyperon resonances in the $t$- and $u$-channel, as shown in
Fig.~\ref{fig:feynman}. 
A long standing problem in the isobar model is the inclusion
of nucleon, kaon, and hyperon resonances. The number of resonances
available from the PDG listing \cite{pdg} 
was considered as too large. In the past, the number of experimental
data was very limited. Using a few well known resonances one could
explain all available data. In fact, the main topic of discussions 
in the past is how to constrain the number of resonances in the
phenomenological models. However, the situation seems to change
with the coming of precise experimental data from recent measurements
in modern laboratories.
These data include not only differential cross sections, but also
single and double polarization observables that are very sensitive
to the choice of resonances in the model. Thus, the new data could
provide a stringent constraint to the existing phenomenological models. 
In fact, the new data can shed more light on the existence of the
missing resonances.

In this study we include all nucleon resonances 
in the PDG listing \cite{pdg} that have
masses between the $K^+\Lambda$ threshold (1.609 GeV) and 2.2 GeV.
The resonance spin is limited only up to 3/2 in order to simplify
the formalism. In addition, we also include the $P_{11}(1840)$ state, 
which was found in Ref.~\cite{Sarantsev:2005tg} to have 
important contribution in the photoproduction of $K^+\Lambda$, $K^+\Sigma^0$, 
and $K^0\Sigma^+$. Our recent study \cite{role_p13} corroborates
this finding. In the $t$-channel, we use two vector mesons with different
parities, i.e. the $K^*(892)$ and $K_1(1270)$, that have been shown
to provide a significant suppression to the divergence of the Born 
terms. As a starting point, we include two hyperon resonances in
the $u$-channel, which were claimed by Ref.~\cite{janssen} to provide
also significant suppression, so that the hadronic form factor cut-off
can be constrained to a reasonable value.
Furthermore, to account for hadronic structure in the hadronic vertices
we include hadronic form factors in a gauge invariant fashion
\cite{Haberzettl:1998eq}. For this purpose, we choose the dipole-type, 
\begin{eqnarray}
  F (q^2) = \frac{\Lambda_{\rm had}^{4}}{\Lambda_{\rm had}^{4}+ 
    \left(q^2-m^{2}\right)^{2}} ,
  \label{eq:dipole}
\end{eqnarray}
where $\Lambda_{\rm had}$ is the cut-off parameter, $q^2$ is the
squared of four-momentum of the off-shell intermediate particle
with the corresponding mass $m$. Note that this form has been
used in most analyses of the meson-nucleon scattering as well
as meson photoproduction. Clearly, the use of other types of
form factor is also possible in this case and the effect of different 
hadronic form factors on the isobar model could become an 
important issue in kaon photoproduction \cite{hadronic-ff}.

\section{Numerical Result}
\label{sec:result}

\subsection{Reanalyzing the Old Models}
As a first step, we try to reproduce the 
claim in Ref.~\cite{janssen}, i.e. certain hyperon resonances
can help to suppress the diverging Born terms.
To this end we fit a model
consisting of the Born terms, $K^*(892)$, $K_1(1270)$, 
$N(1650)S_{11}$, $N(1710)P_{11}$, and $N(1720)P_{13}$
resonances. For the sake of brevity we call this Born
and resonance configuration as the standard configuration.
Using this standard configuration we fit the model to the
SAPHIR \cite{saphir98} and new data and call the result as
Model A and B, respectively. In the next step we add the
missing $D_{13}(1895)$ nucleon resonance \cite{missing-d13} 
into the two models
and repeat the fitting process. Having finished this step
we continue to include the $\Lambda(1800)S_{01}$, 
$\Lambda(1810)P_{01}$ in the standard configuration 
and repeat the fitting process. Finally, we include 
the $D_{13}(1895)$, $\Lambda(1800)S_{01}$, and 
$\Lambda(1810)P_{01}$ in the model and fit to the
SAPHIR and new data as before. The complete result
obtained from this procedure is given in Table~\ref{tab:janssen}
in terms of the obtained $\chi^2/N$, where $N$ indicates the
number of degrees of freedom, i.e. $N=N_{\rm data}-N_{\rm parameter}$.
We also quote the result of Ref.~\cite{janssen} 
in Table~\ref{tab:janssen} for comparison.
Note that in order to maintain consistency with the claim of  
Ref.~\cite{janssen}, in all configurations we limit the
hadronic form factor cut-off to be $\Lambda \geq 1.6$ GeV
during the fitting process.

\begin{table}[h]
  \centering
  \caption{Comparison between $\chi^2/N$ obtained by different number 
    and configuration of resonances. Model A has the same resonance 
    configuration as in Ref.~\cite{janssen} and fitted to the same (SAPHIR) 
    data. Model B uses the same configuration but fitted to the new data.}
  \label{tab:janssen}
  \begin{tabular}[c]{lccc}
    \hline\hline\\[-1ex]
    Configuration & Ref.~\cite{janssen} & Model A & Model B \\[1.5ex]
    \hline\\[-1ex]
    Standard & 10.32 & 8.11 & 14.05\\
    Standard + $N(1895)D_{13}$ & 7.38 & 4.11 & 6.12\\
    Standard + $\Lambda(1800)S_{01}$, $\Lambda(1810)P_{01}$&3.43&4.77&9.18\\
    Standard + $N(1895)D_{13}$, $\Lambda(1800)S_{01}$, 
    $\Lambda(1810)P_{01}$& 2.65& 2.68& 4.07\\[1.5ex]
    \hline\hline
  \end{tabular}
\end{table}

It is obvious that our present calculation is in fair agreement with
the result of Ref.~\cite{janssen}. The difference in the $\chi^2/N$ might
originate from the error-bars used in the fitting process. In our
calculation, we have also included systematic error-bars in addition
to statistical ones. This is reflected by the fact that the
obtained $\chi^2/N$ in Model A is mostly smaller than those in
Ref.~\cite{janssen}. Nevertheless, Model A reveals an interesting 
fact, i.e. the use
of the $D_{13}(1895)$ resonance to suppress the diverging Born terms 
is more effective than the use of two hyperon resonances, the
$\Lambda(1800)S_{01}$ and $\Lambda(1810)P_{01}$, as suggested
by Ref.~\cite{janssen}. Indeed, this phenomenon also appears when 
we use the new experimental data (with more than 3500 data points) 
as shown by Model B in the fourth column of  Table~\ref{tab:janssen}. 

\begin{figure}[t]
\centering
  \includegraphics[width=0.5\textwidth]{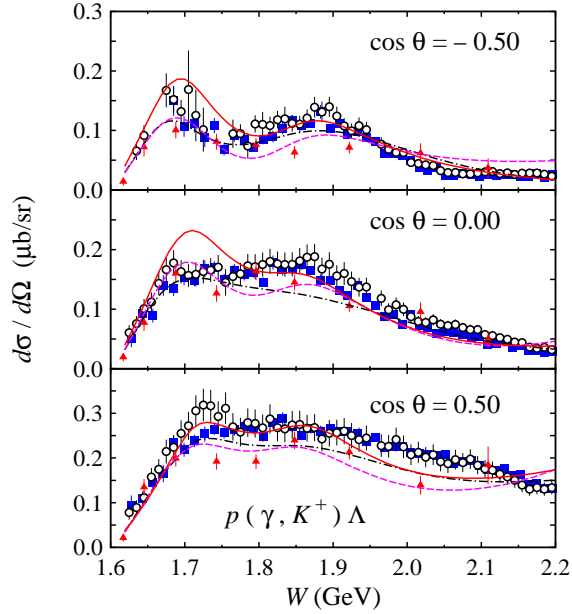}
  \caption{Comparison between differential cross sections 
    obtained from Model A (dashed lines), Model B (solid lines),
    and Kaon-Maid (dash-dotted lines).
    Experimental data are from the 
    CLAS collaboration (solid squares \cite{Bradford:2005pt}
    and open circles \cite{mcCracken}) and SAPHIR collaboration
    (solid triangles) \cite{saphir98}.}
\label{fig:dif1}
\end{figure}

By comparing the results of Model A and B 
in the third and fourth columns of  Table~\ref{tab:janssen}, 
respectively, it is apparent that the resonance configuration
suggested by Table~\ref{tab:janssen} is unable to produce a
reasonable agreement with experimental data. Furthermore, we can
also clearly see that including the missing $D_{13}(1895)$ 
nucleon resonance as suggested in Ref.~\cite{missing-d13}
results in a significant reduction of $\chi^2/N$. 
Nevertheless, combining this resonance with 
$\Lambda(1800)S_{01}$ and $\Lambda(1810)P_{01}$ 
in the standard configuration will reduce the $\chi^2/N$ 
to 4.07, which is still beyond a reasonable value
for a good agreement between model calculation
and experimental data. The reason for that is obviously
displayed in Fig.~\ref{fig:dif1}, where we compare the
calculated differential cross sections obtained from Model A and 
B with the prediction of Kaon-Maid \cite{kaon-maid}.
Presumably, more nucleon resonances are required to explain
the structures, which cannot be reproduced by both Model A
and B. This topic will be discussed in the next
subsection, when we include the higher nucleon resonances, 
which are found to be important in the coupled-channels
study \cite{Sarantsev:2005tg}. We
note that the higher nucleon resonances have been also found to
play an important role in the $K^*\Lambda$ photoproduction \cite{kim}.

\begin{figure}[t]
\centering
  \includegraphics[width=0.8\textwidth]{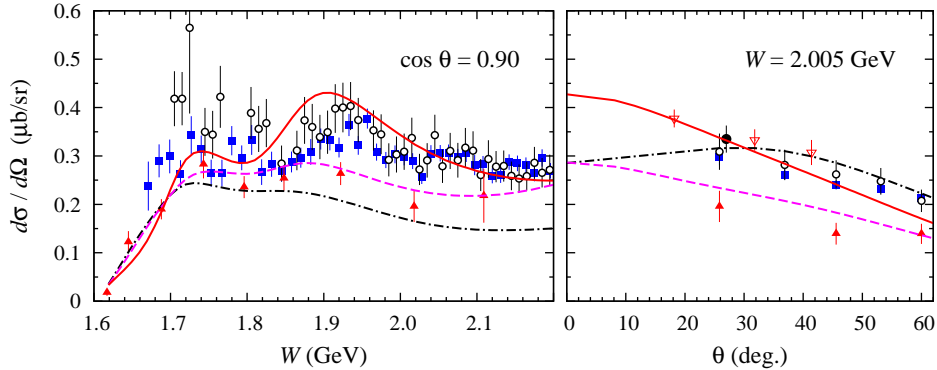}
  \caption{Same as Fig~\ref{fig:dif1}, but for the forward angles, where 
    experimental data are still available for comparison. 
    The behavior of the models at very forward angles can 
    be seen in the right panel. Here, open triangle indicates 
    the LEPS data \cite{sumihama06}, 
    while solid circle comes from an old experiment \cite{fe}. }
\label{fig:dif2}
\end{figure}

At this stage, it is also important to revisit the problem of
kaon photoproduction at forward angles. As discussed in 
Ref.~\cite{Bydzovsky:2006wy}, the use of hadronic form factor 
in Kaon-Maid \cite{kaon-maid} has led to an underprediction of 
differential cross section data at forward angles. Near forward
angles, where experimental data are still available, the situation
is depicted in Fig.~\ref{fig:dif2}. The problem is apparent 
in the case of Kaon-Maid, which originates from the extracted 
form factor cut-off. We notice that the background terms of 
Kaon-Maid is strongly suppressed by a very soft form factor, 
i.e. the dipole form factor given by Eq.~(\ref{eq:dipole}) 
with $\Lambda_{\rm had}=0.64$ GeV. 

However, in the case of Model A and B we 
obtain the relatively larger cut-offs. i.e. $\Lambda_{\rm had}=1.90$ 
GeV and 1.60 GeV, respectively, which correspond to
relatively harder form factors. Thus, both models do not 
exhibit a problem at forward angles as in the case of Kaon-Maid, 
although we also notice that Model A cannot reproduce 
experimental data at this kinematics, since it was fitted to
old SAPHIR data \cite{saphir98}. Nevertheless, from the right
panel of Fig.~\ref{fig:dif2} we can see that the hard 
hadronic form factors in both models are represented by the increase
of the cross section as the kaon angle $\theta$ approaches forward
directions.

We conclude this subsection by emphasizing that to suppress
the diverging Born terms the use of
the $\Lambda(1800)S_{01}$ and $\Lambda(1810)P_{01}$ hyperon
resonances as suggested by Ref.~\cite{janssen}
is less effective than the use of the missing $D_{13}(1895)$ 
nucleon resonance as proposed in Ref.~\cite{missing-d13}.
However, it is true that the soft hadronic cut-off leads
to a strong suppression of the cross section at forward 
directions.

\subsection{Adding More Nucleon Resonances}
From the discussion in the previous subsection it is obvious that
the number of resonances must be increased for the purpose of
explaining 
other structures appearing in the new data and, therefore,
minimizing the $\chi^2$. Using the modified standard configuration 
we obtained the best $\chi^2/N=4.07$ (see Model B, the fourth column 
of Table \ref{tab:janssen}), which indicates a rather poor agreement
between model calculation and experimental data. To this end, we will
use the nucleon resonance configuration in the 
isobar model developed in our recent study \cite{role_p13},
which is able to nicely reproduce the presently available data. 
The model is constructed by using the same standard configuration
(see Table \ref{tab:janssen}) and, in addition, the 
$P_{13}(1720)$, $P_{11}(1840)$, $P_{13}(1900)$, 
$D_{13}(2080)$, $S_{11}(2090)$, as well as  $P_{11}(2100)$ resonances.
Note that these resonances
have been found to be important in photoproduction of 
$K^+\Lambda$, $K^+\Sigma^0$, and $K^0\Sigma^+$ \cite{Sarantsev:2005tg}. 
In the original model \cite{role_p13}, the $\Lambda^*(1800)$ and 
$\Lambda^*(1810)$ hyperon resonances have been also included in order
to achieve the best agreement with experimental data. In this
study we start our fitting process by including all available
$\Lambda^*$ and $\Sigma^*$ in the PDG listing \cite{pdg} with spin 1/2.
We do not include higher spin hyperon resonances for the
sake of simplicity. Table \ref{tab:resonance} summarizes the nucleon 
and hyperon resonances used in the present study.

\begin{table}[tb]
  \centering
  \caption{Nucleon and hyperon resonances used in the present investigation
    along with their properties. Most of the data are taken from Particle
    Data Group (PDG) listings \cite{pdg}. Otherwise, the values originates 
    from our previous calculation \cite{role_p13}.}
  \label{tab:resonance}
  \begin{tabular}[c]{cccc}
    \hline\hline\\[-1.2ex]
    \multicolumn{2}{c}{Resonance} & Mass & Width \\[1ex]
    \cline{1-2} & \\
    Short Symbol & PDG Symbol & (MeV) & (MeV) \\[1.2ex]
    \hline\\[-1ex]
    $S_{11}(1650)$ & $N(1650)S_{11}$ & 1650 & 150 \\
    $S_{11}(2090)$ & $N(2090)S_{11}$ & 2090 & 400 \\
    $P_{11}(1710)$ & $N(1710)P_{11}$ & 1710 & 100 \\
    $P_{11}(1880)$ & $N(1880)P_{11}$ & 1952 & 413 \\
    $P_{11}(2100)$ & $N(2100)P_{11}$ & 2100 & 113 \\
    $P_{13}(1720)$ & $N(1720)P_{13}$ & 1720& 150 \\
    $P_{13}(1900)$ & $N(1900)P_{13}$ & 1900& 498 \\
    $D_{13}(1700)$ & $N(1700)D_{13}$ & 1700& 100 \\
    $D_{13}(2080)$ & $N(2080)D_{13}$ & 2080& 450 \\
    $\Lambda^*(1405)$ & $\Lambda(1405)S_{01}$  & 1406 & 50 \\
    $\Lambda^*(1600)$ & $\Lambda(1600)P_{01}$  & 1600 & 150 \\
    $\Lambda^*(1670)$ & $\Lambda(1670)S_{01}$  & 1670 & 35 \\
    $\Lambda^*(1800)$ & $\Lambda(1800)S_{01}$  & 1800 & 300 \\
    $\Lambda^*(1810)$ & $\Lambda(1810)P_{01}$  & 1810 & 150 \\
    $\Sigma^*(1660)$  & $\Sigma(1660)P_{11}$   & 1660 & 100 \\
    $\Sigma^*(1750)$  & $\Sigma(1750)S_{11}$   & 1750 &90  \\[1.5ex]
  \hline\hline
  \end{tabular}
\end{table}

\begin{figure}[t]
\centering
  \includegraphics[width=0.7\textwidth]{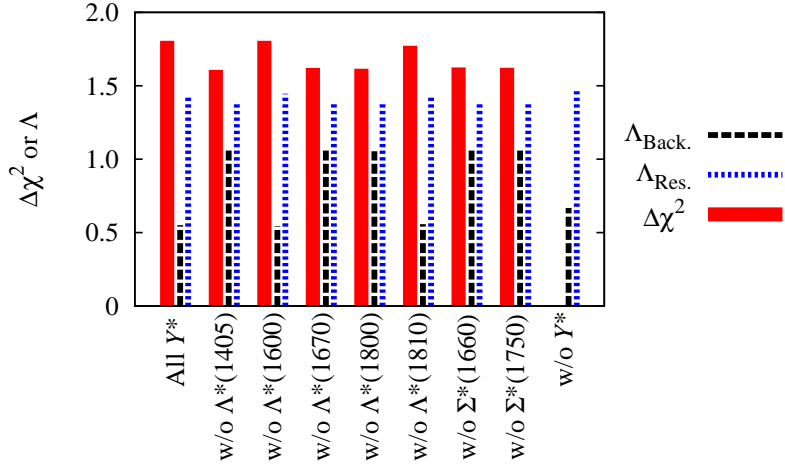}
  \caption{Contribution of the hyperon resonance(s) in reducing
    the $\chi^2$ compared with the extracted background and resonance 
    hadronic form factor cut-offs. Shown in this figure are results 
    from different fits obtained
    with all resonances, without (w/o) specific resonance, and
    without all hyperon resonances included in the model.}
\label{fig:chi2}
\end{figure}

Having finished the fitting process we start excluding each of 
the hyperons and repeat the process in order to estimate the
role of each hyperon in minimizing the $\chi^2/N$ and, at the same
time, keeping the hadronic form factor cut-off always reasonably
hard. The result is shown in Fig.~\ref{fig:chi2}, where we have
defined
\begin{eqnarray}
  \label{eq:delta-chi}
  \Delta\chi^2 = \frac{|\chi^2_{\rm excl.}-\chi^2_{\rm no}|}{1000} ,
\end{eqnarray}
with $\chi^2_{\rm excl.}$ indicates the $\chi^2$ obtained 
with one particular hyperon excluded from the model, whereas 
$\chi^2_{\rm no}$ is the $\chi^2$ obtained without including
the hyperon resonance at all. The factor of 1000 in the denominator
is used in order to fit the result on the scale.
Thus, the $\Delta\chi^2$ in 
Eq.~(\ref{eq:delta-chi}) measures the role of one particular hyperon
resonance in reducing the $\chi^2$, i.e. the larger the value 
of $\Delta\chi^2$, the more important is the resonance. 

From Fig.~\ref{fig:chi2} we can clearly see that only two hyperon
resonances can significantly reduce the 
$\chi^2$, i.e. the $\Lambda^*(1600)$ and $\Lambda^*(1810)$. 
The latter has been found as an important hyperon resonance in
kaon photoproduction since more than a decade ago, but the
former is new. Interestingly, excluding these resonances from
the model leads to very soft background cut-offs,
which is clearly exhibited by the shorter dashed (black) lines in
Fig.~\ref{fig:chi2}. Therefore, both hyperon resonances
are important to reduce the $\chi^2$ as well as to keep
the hadronic cut-off reasonably hard. 

\begin{figure}[t]
\centering
  \includegraphics[width=0.8\textwidth]{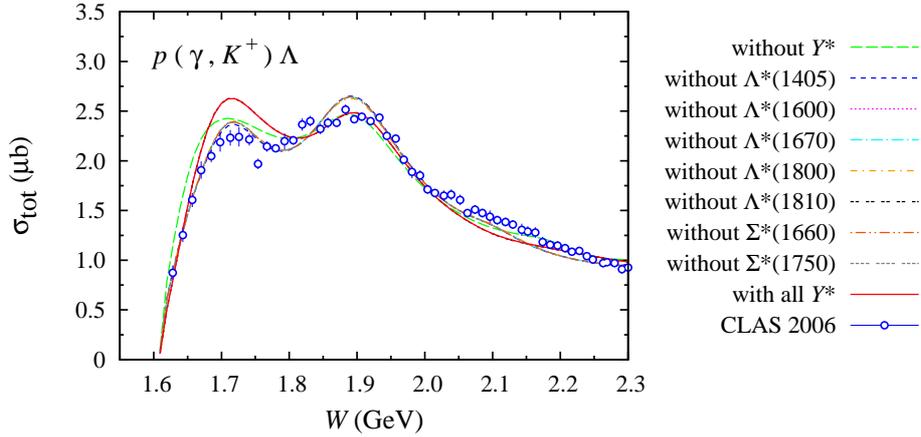}
  \caption{Total cross section of the $K^+\Lambda$ photoproduction
    obtained from a number of model calculations. Notation of the
    curves are given on the right hand side of the figure. 
    Experimental data are from the CLAS collaboration 
    \cite{Bradford:2005pt}. These data were not used in the
    fitting process of the present investigation.}
\label{fig:kltot}
\end{figure}

It is also important to note that in the resonance sector
there is no need to suppress the amplitude, since the resonances
do not produce a diverging amplitude, in spite of the fact that
the covariant Feynman diagrams also create additional background
amplitudes. 
Furthermore,
Fig.~\ref{fig:chi2} implies that including all hyperon
resonances in the model were not recommended, since the background
cut-off would significantly drop to a very soft value. 
This indicates that the $\chi^2$ alone cannot be used
as the only measure for a good agreement with experimental data.
The soft hadronic form factor would be also obtained if we 
did not include any hyperon 
resonance.
To conclude this result, we may say that 
Fig.~\ref{fig:chi2} reveals the dual-role
of hyperon resonances, i.e. reducing the $\chi^2$ value and,
simultaneously, keeping
the hadronic cut-off to a reasonably hard value.

Comparison between experimental data and model 
calculations with or without (a specific or all)
hyperon resonances are displayed in 
Fig.~\ref{fig:kltot}. Here we can see that the model obtained 
by including all hyperon resonances and that obtained by excluding
the $\Lambda^*(1600)$ or $\Lambda^*(1810)$ hyperon resonance 
fail to reproduce the first peak of the total cross section.
Since a number of curves in this figure is coincidence with each other,
we have checked the numerical results and confirmed this
finding. 
Again, this indicates that we cannot neglect these hyperon
resonances in $K^+\Lambda$ photoproduction. 

\begin{figure}[t]
\centering
  \includegraphics[width=0.8\textwidth]{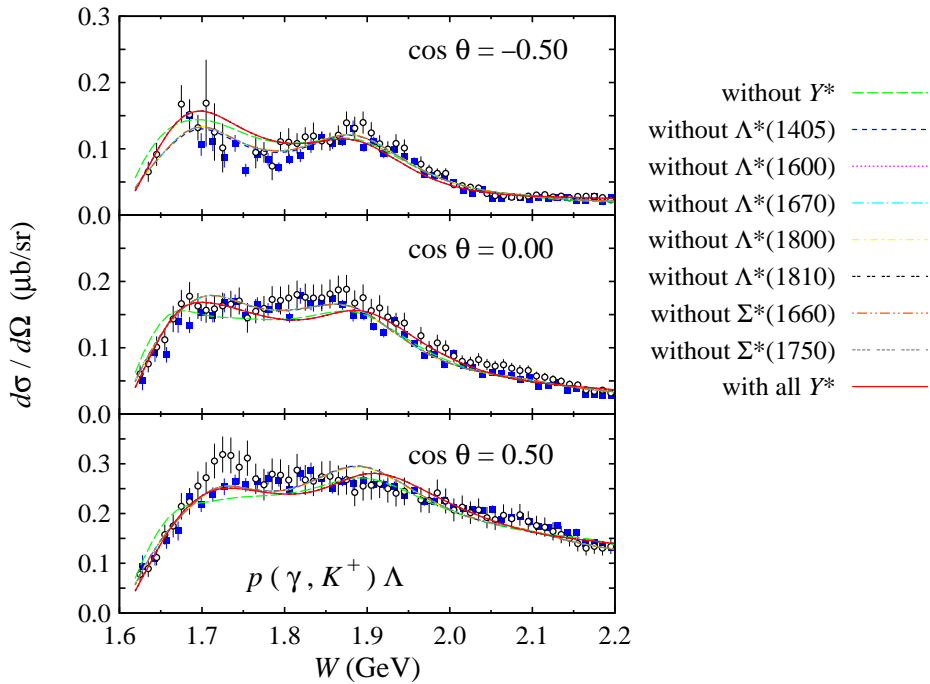}
  \caption{Same as Fig~\ref{fig:kltot}, but for differential
    cross section. Experimental data are taken from the 
      CLAS collaboration (solid squares \cite{Bradford:2005pt}
      and open circles \cite{mcCracken}).}
\label{fig:dkpl_w}
\end{figure}

For the second peak in the total cross section 
we see that the predictions of these three
models seem to overpredict the data. However, we observe that
this happens because at this point 
the new version of CLAS differential cross section 
data  \cite{mcCracken}
are substantially larger than the previous 
version  \cite{Bradford:2005pt}. 
This could be
the origin of the problem. The comparison in this
case is given in Fig.~\ref{fig:dkpl_w} for the differential
cross section.

\begin{figure}[h]
\centering
  \includegraphics[width=0.8\textwidth]{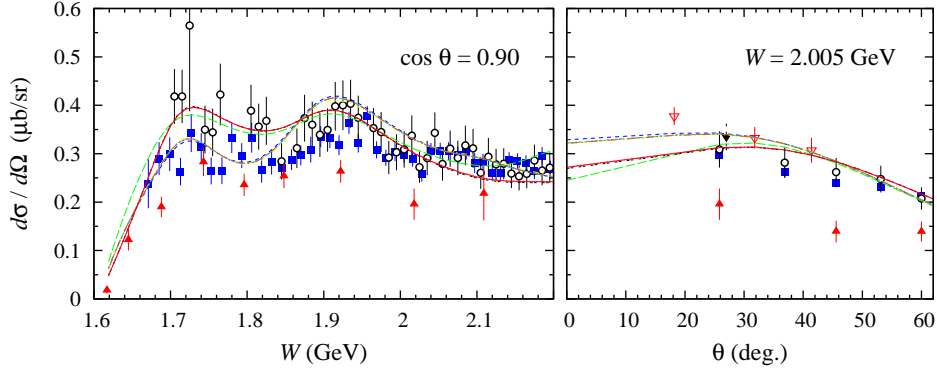}
  \caption{Differential cross sections in the forward region
    as in Fig.~\ref{fig:dif2}, but for different fits
    using different hyperon resonances. Notation of the curves
    is as in Fig.~\ref{fig:kltot}.}
  \label{fig:forward1}
\end{figure}

The more obvious difference between model calculations is
observed in the forward region, as depicted in Fig.~\ref{fig:forward1}.
From the left panel of Fig.~\ref{fig:forward1} we might conclude 
that 
the problem in the fitting process is due to the scattered 
experimental data.
However, the right panel of Fig.~\ref{fig:forward1} displays
the important message: using the soft hadronic form factor
obtained in a model with all hyperon resonances included, 
a model without
$\Lambda^*(1600)$, or a model without $\Lambda^*(1810)$, 
as shown in Fig.~\ref{fig:chi2}, results in a suppressed 
differential cross section at very forward angles.
Therefore, the result of our present study provides a more solid 
evidence of the origin of this suppression, which was predicted 
previously in Ref.~\cite{Bydzovsky:2006wy}.

\begin{figure}[t]
\centering
  \includegraphics[width=0.5\textwidth]{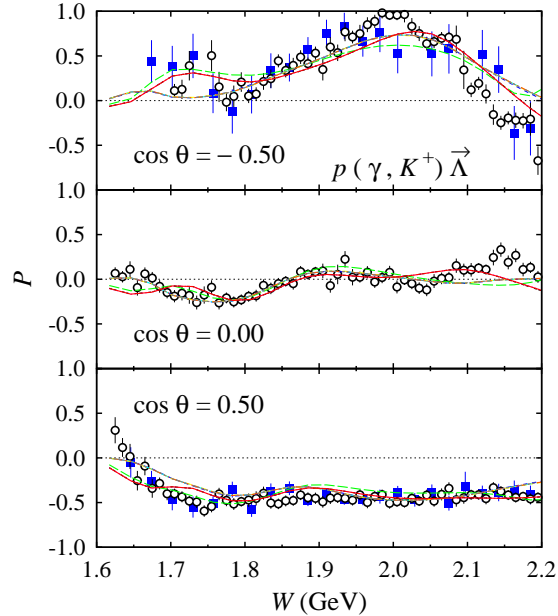}
  \caption{Comparison between model calculations and experimental data
    for the recoil polarization observable $P$. Experimental data 
    are taken from the CLAS collaboration (solid squares 
    \cite{Bradford:2005pt} and open circles \cite{mcCracken}).
    Notation of the curves is as in Fig.~\ref{fig:kltot}.}
\label{fig:pollam}
\end{figure}

\begin{figure}[t]
\centering
  \includegraphics[width=0.6\textwidth]{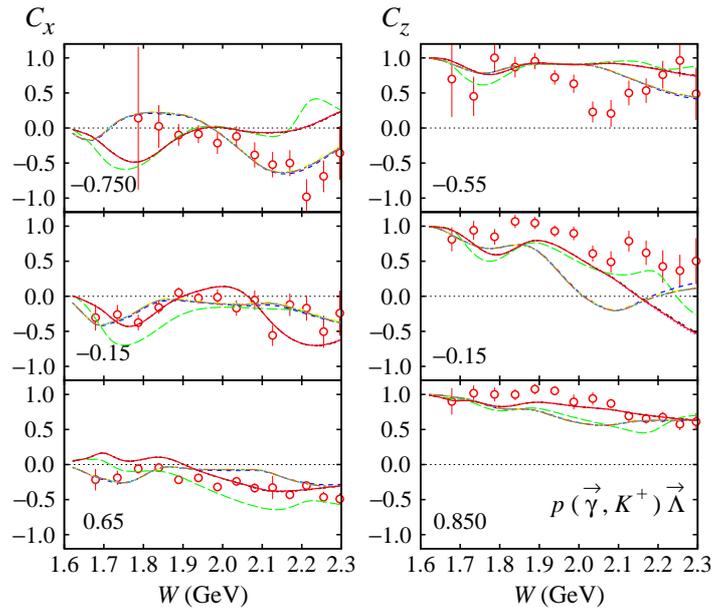}
  \caption{The beam-recoil double polarization observables $C_x$ 
    (left panel) and $C_z$ (right panel). Experimental data are from the 
    CLAS collaboration \cite{Bradford:2007}.
    Notation of the curves is as in Fig.~\ref{fig:kltot}.}
\label{fig:cx_cz}
\end{figure}

It has been also widely known that both single- and 
double-polarization observables are sensitive to the choice
of resonances in the model. Therefore, in Figs.~\ref{fig:pollam}
and \ref{fig:cx_cz} we display the result of our present 
investigation in the case of recoiled hyperon polarization $P$
and beam-recoil double polarizations $C_x$ and $C_z$, respectively.
However, we observe that the recoil polarization does not show
a remarkable sensitivity, except at $W\leq 1.7$ GeV. In this
kinematics we find sizable differences between models, but
within the present error bars they are still difficult to distinguish.

The beam-recoil double polarization $C_x$ 
shown in Fig.~\ref{fig:cx_cz} is 
found as the most sensitive observable for our
present purpose, especially at backward directions.
In  Fig.~\ref{fig:cx_cz} it is shown that excluding
the $\Lambda^*(1600)$ and $\Lambda^*(1810)$ resonances
results in a clear disagreement between data and 
model calculations. We notice that the difference
decreases in the forward region. In the case of
double polarization $C_x$ the situation is difficult,
due to the deficiency of the model, except at a very
forward angle. Unfortunately, in this kinematics
the difference between model calculations is
smaller than that obtained in other kinematics. Thus, this observable
is not recommended for the present purpose.

\begin{figure}[h]
\centering
  \includegraphics[width=0.8\textwidth]{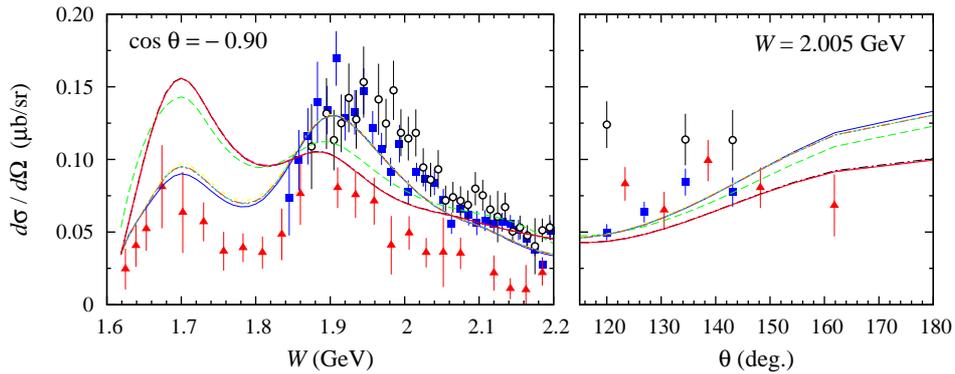}
  \caption{Differential cross sections as in 
    Fig.~\ref{fig:forward1}, but for the backward directions.}
  \label{fig:backward}
\end{figure}

Our present finding is therefore partly different from
the result of the previous work \cite{janssen}, i.e. instead
of the $\Lambda(1800)S_{01}$ resonance we found the 
$\Lambda(1600)P_{01}$ resonance to be the important one.
This difference presumably originates from the new data used
in the present study.

From the propagator behavior it could be expected that the 
$u$-channel resonances would have dominant effects in the 
backward directions, in Fig.~\ref{fig:backward}
we display our result for this kinematics. Obviously the effect
of using different hyperon resonances is apparent here. 
By comparing Figs.~\ref{fig:forward1} and \ref{fig:backward} 
we can see that the effect found in the 
forward region is even amplified in the backward direction.
The left panel of Fig.~\ref{fig:backward} clearly confirms 
that the effect is largest at kaon angle 180$^\circ$.
Unfortunately, experimental data at this kinematics cannot
resolve this effect.

\subsection{Sensitivity of the Result to the Hadronic
  Form Factor Cut-Off}
\begin{figure}[t]
\centering
  \includegraphics[width=0.45\textwidth]{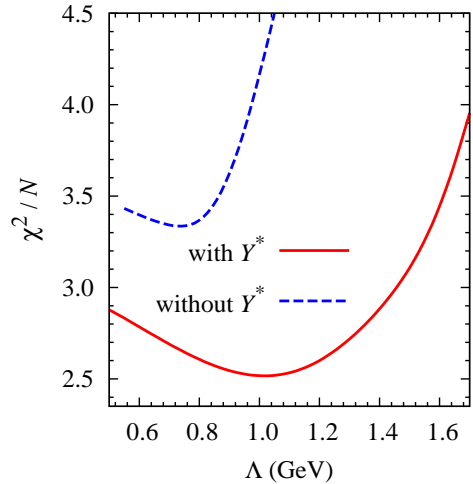}
  \caption{Comparison between the $\chi^2/N$ obtained 
    from a model including hyperon resonances (solid line) and that 
    excluding hyperon resonances
  (dashed line) plotted as a function of the hadronic form factor
  cut-off of the Born terms.}
\label{fig:cut-off}
\end{figure}

How sensitive our result to the variation of the hadronic form factor
cut-off would be also an interesting question to address. To 
investigate this sensitivity we make 
use of the isobar model proposed in Ref.~\cite{role_p13}, 
where the $\Lambda(1800)S_{01}$ and $\Lambda(1810)P_{01}$ hyperon
resonances have been already included. However, to account for
the result obtained in the previous section, in the present discussion
we add the $\Lambda(1600)P_{01}$ hyperon resonance to the model 
and refit the model predictions to 
experimental data. The $\chi^2/N$ reduces from 2.57 
to 2.49. The hadronic form factor cut-offs for the background and
resonance terms are obtained to be 1.055 and 1.385 GeV, respectively.
Thus, we believe that the model is sufficiently good for our
present purpose.
Using this model we vary the background cut-off from 0.5 to 2 GeV
with 0.05 GeV step, refit the experimental data, and calculate 
the $\chi^2/N$. The result is
depicted in Fig.~\ref{fig:cut-off} by the solid line.
For comparison, we also propose a model without hyperon resonance,
for which the obtained hadronic cut-off is 0.769 GeV and
the best $\chi^2/N$ is 3.34.
The corresponding variation of $\chi^2/N$ as a function 
of the background cut-off 
is shown in Fig.~\ref{fig:cut-off} by using the dashed line.

It is obvious that the three hyperon resonances are able to significantly
improve the agreement with experimental data and shift the desired
cut-off to a larger value. Although the two cases are clearly different,
both share the same behavior, i.e. increasing the cut-off value would
directly increase the disagreement between model calculations 
and experimental data. Especially, in the case of the model without
hyperon resonance, where the $\chi^2/N$ increases dramatically
with slightly increasing the cut-off. The result indicates that
in these models obtaining a softer hadronic form factor would be
much easier than a harder one. Put in other words, the models
are less sensitive to the variation of the cut-off in the 
smaller cut-off region, but very sensitive in the larger one.

\section{Summary and Conclusion}
\label{sec:conclusion}
We have investigated the role of hyperon resonances in 
the elementary photoproduction of  $K^+\Lambda$ 
by using isobar models. To this end, we have reproduced the
previous 
claim that the $\Lambda(1800)S_{01}$ and $\Lambda(1810)P_{01}$
could increase the hadronic form factor cut-off to a harder
region, by using the SAPHIR 1998 data. Using the new data, we
have shown that the resonance configuration used to reproduce 
the SAPHIR 1998 data is insufficient to achieve a
reasonably good agreement between model calculations and data.
To overcome this problem we have added more nucleon resonances 
in the model. The model is  able to nicely reproduce the 
new data. By making use of this model we have
investigated all spin 1/2 hyperon resonances listed by the
PDG. It is found that the $\Lambda(1600)P_{01}$ and $\Lambda(1810)P_{01}$
resonances are required to reduce the $\chi^2$ and, simultaneously,
to keep the hadronic form factor reasonably hard. Excluding one of 
these (or both) resonances from the model results in a suppression 
of differential cross section at forward angles. Our finding is therefore 
partly different from the result of previous studies. Our result 
confirms that the effect of the hyperon  resonances is largest in 
the backward direction. The obtained model is
very sensitive to the variation of the hadronic cut-off for 
the background terms, especially if no hyperon
resonance were included. The sensitivity of the model implies
that obtaining a model with 
softer cut-off is much easier than that with harder one.

\begin{acknowledgements}

TM acknowledges the supports
from the University of Indonesia
and the Competence Grant of the Indonesian 
Ministry of National Education are gratefully
acknowledged.

\end{acknowledgements}

\end{document}